\documentclass{article}
\usepackage{float}
\usepackage[english]{babel}
\usepackage{authblk}
\usepackage{mhchem}

\usepackage[a4paper,top=2cm,bottom=2cm,left=3cm,right=3cm,marginparwidth=1.75cm]{geometry}

\usepackage{amsmath}
\usepackage{graphicx}
\usepackage[colorlinks=true, allcolors=blue]{hyperref}
\usepackage{subfigure}
\usepackage{subcaption}
\usepackage{caption}

\title{Insights on molecular \ce{P} implantation for scalable spin-qubit arrays}
\author[]{Tomás Fernández Bouvier}
\author[]{Ville Jantunen}
\author[]{Saana Vihuri}
\author[]{Alvaro López Cazalilla}
\author[]{Flyura Djurabekova}
\affil{Accelerator Laboratory, University of Helsinki}

\begin{document}
\maketitle

\begin{abstract}
Quantum information technologies hold immense promise, with quantum computers poised to revolutionize problem-solving capabilities. Among the leading contenders are solid-state spin-qubits, particularly those utilizing the spin of phosphorous donors ($^{31}P$). While significant progress has been made in enhancing quantum coherence and qubit control, challenges persist, notably in achieving precise and scalable \ce{P} placement in Si substrate. This paper investigates by means of molecular dynamics the use of molecular \ce{PF2} ions for implantation, aiming to reduce placement uncertainty while maintaining detection efficiency. We examine energy transfer, molecule integrity, implantation profiles, electronic signal components, and stable damage. Among other things we find that the assumption that the molecule only breaks apart immediately due to the presence of an a-SiO$_{2}$ layer on the surface of the crystal and that the intensity of the electronic signal from ion-solid interactions does not correlate necessarily with the penetration depth of P. 
\end{abstract}

\section{Introduction}

Among the favourite candidates for solid-state spin qubits is the spin of the outermost electron of a phosphorous donor ($^{31}P$) \cite{Kane1998}. Since its conception the low spin-medium interaction was improved by successive corrections to achieve currently unprecedentedly high spin coherence times \cite{coher1, coher2}. 

The preferred substrate for manufacturing \ce{P} spin-donor qubits today is silicon \cite{Morton2011} because of its abundance, well-developed Si-based semiconductor industry 
and a low spin-orbit coupling \cite{randall2009atomic,STML2}.
For optimal operation of the qubit, effective control and readout, the placement of the donor must not exceed 7-20 nm beneath the surface \cite{location_implant, elstop_detection}.
Two techniques are used for the precise placement of the qubits in the substrate. On them is based on scanning tunnelling microscope (STM) lithography, which achieves high placement precision but requires a subsequent epitaxial substrate regrowth. Moreover, the donor is introduced 
not as a single atom but as a cluster\cite{Büch2013}. On the other hand, the modern ion-beam technologies allow to implant \ce{P} atoms one by one, circumventing the regrowth process. Ion beams are also a standard technique commonly employed in the semiconductor industry\cite{van2015single, detection_confidence} and, hence, can be immediately available for solid state qubit production without significant development.

 The donor placement via ion implantation, however, is to date orders of magnitude more uncertain than in the STM lithography. Luckily, during the process, the ion interacts with the solid through electronic stopping creating a signal which enables the determination of the final position of the implanted atom\cite{detection_confidence}. The intensity of the signal correlates with the energy and the atomic number of the ion. The above-mentioned donor placement depth window corresponds to the maximum of the 9 keV \ce{P} ion range.  
 The higher ion energies will result in higher detection efficiency, but also  in stronger damage and, hence, higher uncertainty in placement. Therefore, there is a trade-off between the placement precision and the detection efficiency. Recently, it was demonstrated that using molecular \ce{PF2} ions could be a solution, since they create a high signal while keeping the placement uncertainty close to the 9 keV \ce{P} conditions \cite{jakob2023scalable}.

As computationally very efficient, Binary Collision Approximation (BCA)  methods are commonly used to estimate the ion range in materials, also for the molecular ions, assuming that a molecule immediately breaks apart right after the impact with the target \cite{SRIM}. If so, a BCA method can, in general, provide sufficient insight on depth and lateral resolution for the positioning of the implanted donor \cite{holmes2023improved}.
However, there are some processes that remain hidden to the BCA approach, which does not include many-body effects \cite{nordlund2019273}. For a molecular implantation, an accurate description of the total electronic signal recorded during the experiment, and how this signal can be used to detect channeling events need to be clarified. Finally, there is a lack of understanding of the radiation damage in the Si lattice caused by \ce{PF2} ion implantation compared to the one caused by it's components. Any damage needs to be recovered in a subsequent annealing to guarantee that the implanted donor is electrically activated \cite{Saito1998, Sedgwick_1983}. 

In the present work, we address these questions using molecular dynamics (MD) simulations. We begin by describing the energy loss and transfer due to the presence of an amorphous \ce{SiO2} layer (a-\ce{SiO2}) on the surface of the crystalline Si wafer (c-\ce{Si}), the integrity of the molecule as a function of the depth and how this will lead to overlapping cascades. Then, we discuss the relevance of the use of molecular ions on resulting implantation profiles. Next, we study the different components that are added up in the electronic signal and how the intensity of the latter correlates with the implantation depth and channeling events. Finally, we study the extent of the damage caused by the molecular ions as well as the defects evolution during the post-implantation annealing. 

\section{Simulation methods}

In the present study we used the Molecular Dynamics (MD) method to analyse the difference between a mono-atomic and a molecular ion irradiation of Si. All simulations were performed using the Large-scale Atomic/Molecular Massively Parallel Simulator (LAMMPS) \cite{LAMMPS}.

The fairly large size 43.5$\times$43.5$\times$65.7 nm$^{3}$ of the simulation cell with the crystalline Si structure provided sufficient bulk volume to follow the penetration dynamics of the impacting energetic ions including those moving in the channeling directions. The interaction between Si atoms in our system was described by the Stillinger-Weber potential \cite{Sillinger_weber} (SW). The choice of the potential is justified by the similarity of the cascade evolution predictions by SW and the machine-learning Gaussian Approximation potential (GAP) \cite{hamedani2021}, while the computational efficiency of the former is much higher, which is essential for the purpose of the current study. The energy of the input lattices was first minimised 
using the conjugate gradient algorithm with an extra degree of freedom for the cell size. Then, the system was thermally equilibrated at 300 K using the isothermal-isobaric (NPT) ensemble with the temperature and pressure constants of 0.3 ps and 9.0 ps, respectively, for 100 ps of the  simulated time. After that, the periodic boundary conditions were removed in the Z direction, the bottom 2 nm layer of atoms was frozen to avoid any lattice drift and the system was relaxed in the canonical (NVT) ensemble at 300 K for another 100 ps.

Interaction of ions with the lattice was described using the purely repulsive ZBL potential \cite{zbl}. For the interactions between the atoms within the molecule, we used the Lennard-Jones potential:
\begin{equation}
V_{LJ}=4 \epsilon \left( (\sigma / r)^{12} - (\sigma /r)^{6} \right)     
\end{equation}
 
 where $\epsilon$ = 4.6, 5.078 and 1.6585, while $\sigma$ = 1.408, 2.21 and 2.131 Å for P-F, P-P and F-F interactions, respectively \cite{PF2_params}. For consistency of the simulated results, the initial velocities of the ions were set in such a way that the energy of the \ce{P} atom was 9 keV. This means that the molecular ion \ce{PF2} had the energy of 20.041 keV. Similarly to the experimental condition described in \cite{holmes2023improved}, an incidence angle of 7 degrees from the normal was used to decrease the probability of ion channeling.

The ions were introduced at approximately 1 nm above the surface where interactions of the ion with the lattice are negligible. The lateral position of the ion was randomised within a 0.29 nm$^{2}$ (i.e. 1$\times$1 unit cell) on the XY plane. In the case of molecular ion, the molecule was randomly rotated before the impact. 

\begin{figure}[H]
    \centering
    \subfigure[]{\includegraphics[width=0.42\linewidth]{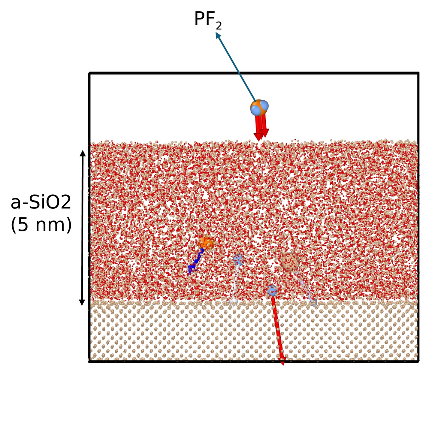}}
    \subfigure[]{\includegraphics[width=0.57\linewidth]{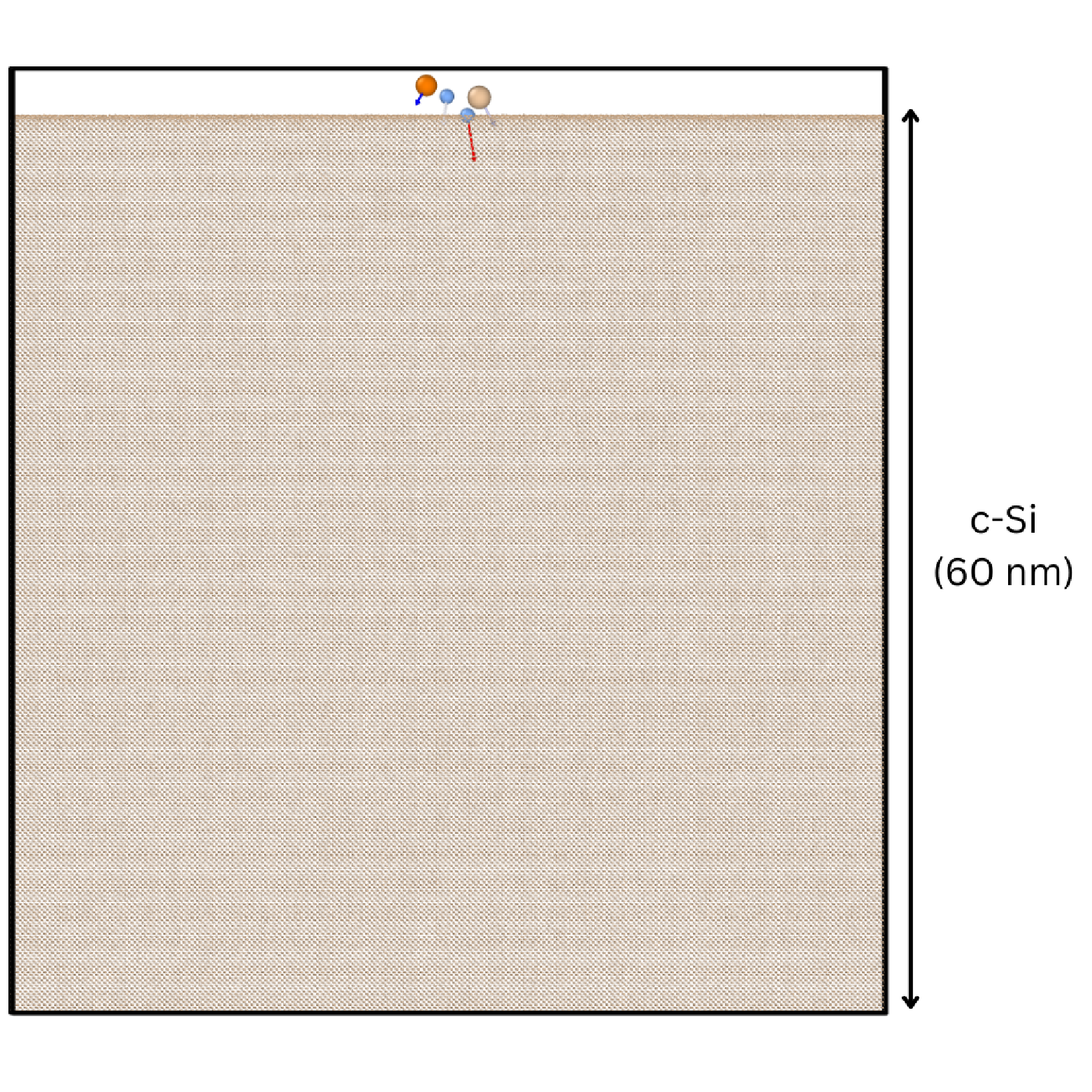}}
    \caption{Simulation setup. a) Small simulation cell containing a 5 nm amorphous silica (a-SiO$_{2}$). High energetic particles are recorded at a-SiO$_{2}$/Si interface. b) Crystalline silicon (c-Si) big cell. Projectiles can either be sampled 1 nm above the surface or be parsed by simulations in the smaller box.}
    \label{fig:setup}
\end{figure}

We also note that there is usually an amorphous silica layer (a-\ce{SiO2}) that grows naturally on the wafer in experiments. This will alter the energy and angle of the incidence, when the ions reach Si wafer. For a molecular ion, the a-\ce{SiO2} layer may affect the integrity of the molecule before it reaches the crystal. 
For comparison, a set of simulations were carried out by sampling the molecule location directly over the pristine c-Si lattice while another set sampled the molecular ion impact on the c-Si surface covered with a 5 nm a-\ce{SiO2} layer.
In the latter simulations, the size of the cell was smaller with the purpose to follow the angular and energy spread of ions entering the Si bulk (see Figure \ref{fig:setup} a)). The interatomic interactions in the latter box were described by using the Tersoff-type potential \cite{MUNETOH2007334}, while for short-distance interactions we used the same ZBL potential. The position and momentum of any particle with the energy at the interface above 2 keV were recorded exactly at the time when the first energetic particle arrives at the a-\ce{SiO2}/c-Si interface. The energetic atoms were then exported to the larger cell of the crystalline Si, see Figure \ref{fig:setup}b), where the simulation was continued. 

For consistency of our simulations, we continued the large scale simulations in the same SW potential, which was preferred over the Tersoff potential because of the reasons mentioned above:
the irradiation damage production in a recently developed GAP \cite{hamedani2021} was qualitatively and quantitatively closer to the SW potential than that of the Tersoff. 
By simulating the interface and the pure c-Si structures separately, we were able to keep the same potential for both cases, enabling comparisons between the resulting damages. 

A border cooling was applied to the 0.2 nm layer using the Berendsen thermostat at 300 K with the temperature constant $\tau = 0.1$ ps in the x and y periodic directions as well as at bottom of the cell in the z direction. An additional frozen layer of 2 Å was added below the latter border to prevent the cell from drifting due to pressure waves. The interior atoms were evolved under the micro-canonical ensemble (NVE). Electronic stopping was added to the atoms surrounded by not less than 3 neighbours, when their energy exceeded 10 eV. The simulations were carried out for 15 ps, time at which most of the damage created is assumed to be stable in MD time-scales. 

Visualisations of the atomic structures, implantation profiles, defect analyses using the Wigner-Seitz method as well as the cluster analyses were performed using the visualisation tool OVITO \cite{ovito}.

\section{Results and Discussion}

\subsection{Molecular versus mono-atomic ion \ce{P} implantation in silicon with and without silica surface layer} 
\begin{figure}[H]
    \centering
    \subfigure[]{\includegraphics[width=0.49\textwidth]{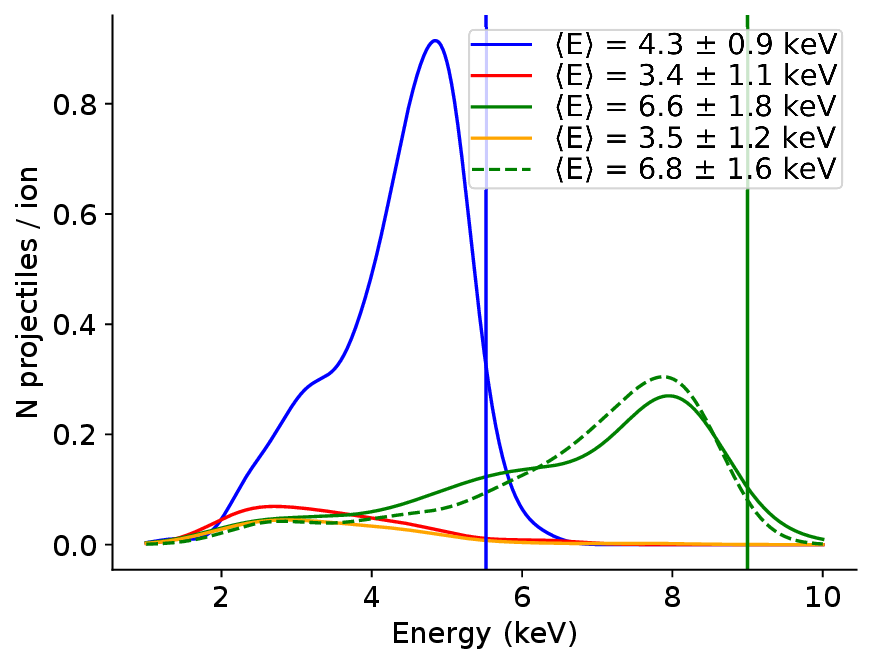}}
    \subfigure[]{\includegraphics[width=0.49\textwidth]{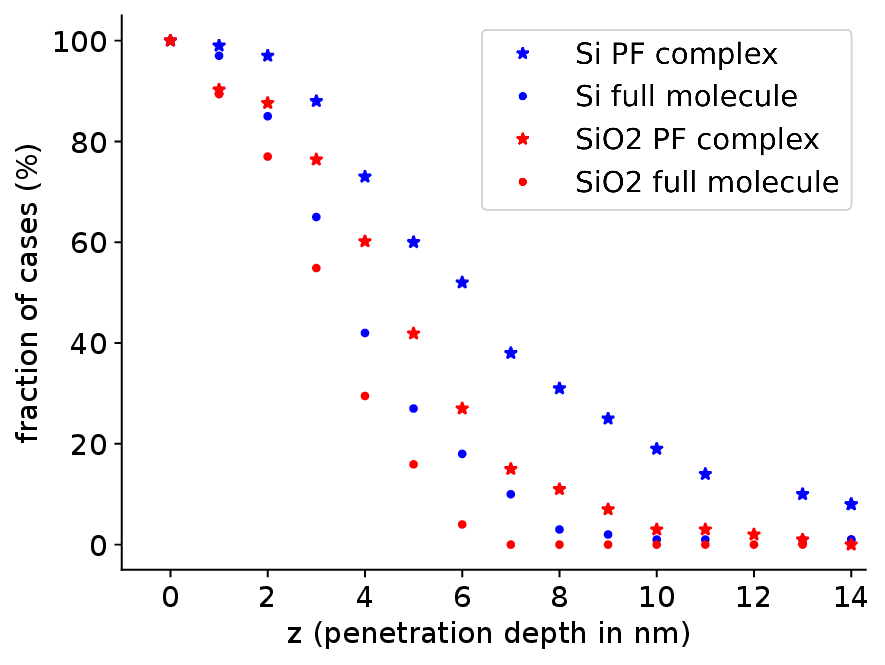}}
    \subfigure[]{\includegraphics[width=\linewidth]{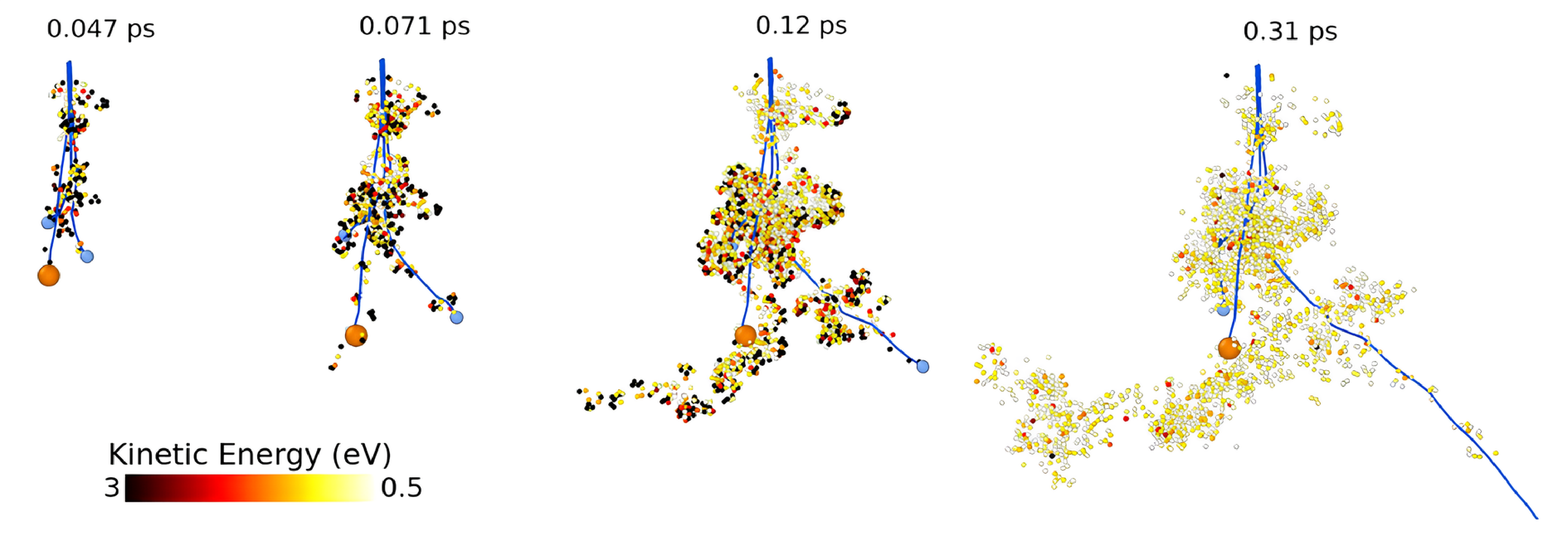}}
    \caption{a) Energy distribution of the particles with an energy higher than 2 keV recorded at the interface between a-SiO$_{2}$ and c-Si for \ce{P} (green), F (blue), O (red) and Si (yellow) when using a molecule (solid) or \ce{P} alone (dashed). Vertical lines correspond to the energy values before impact. b) Number of cases where one or two F lies within 1 nm of the \ce{P} as a function of depth implying a high overlap of their cascades. c) Snapshots from the evolution of a representative implantation case. The P and F atoms are highlighted in orange and purple respectively. From 2D projection it is not clear but there is an overlap of the trajectories (blue trace) followed by the ions.}
    \label{fig:effect_SiO2}
\end{figure}

To analyze the effect of the presence of natural a-\ce{SiO2} layer on top of the c-\ce{Si} substrate on the value of ion energy, when the ions eventually reach Si, we performed 100 simulation runs of the molecular ion implantation into the thin 5 nm a-\ce{SiO2} slab. For each run, the initial position of the center of mass of the molecule and the ion velocities were the same, while orientation of the molecule with respect to the substrate as well as the lateral position were randomised. 

We recorded the energy of the implanted ions at the exit from the slab (i.e. of those which will cross the a-\ce{SiO2}/c-\ce{Si} interface in experiment) and plotted the energy distributions for each species separately in Figure \ref{fig:effect_SiO2}a.
Since the cascade triggered within the amorphous layer may set into motion not only the implanted ions, but also the atoms of the target material, Figure \ref{fig:effect_SiO2}a shows also the energy distributions for all energetic particles that exited the a-\ce{SiO2} slab, i.e. crossing the a-\ce{SiO2}/c-Si interface. Here, green and blue colors show the energy spread for \ce{P} and F ions,  respectively, while the red and yellow colors are used for O and Si recoil atoms, respectively. The numbers in the legend are the mean ion energy that the ion of the corresponding type has at the exit from the a-\ce{SiO2} slab. Although the number of the Si and O recoils is low (see the red and yellow curves in Figure \ref{fig:effect_SiO2}a), the average energy that is brought in by this ions is non-negligible and may create additional damage in the Si layer beneath the a-\ce{SiO2}, also promoting the O diffusion into the c-\ce{Si} bulk. The data was smoothed by applying a kernel density estimator (KDE) fitted to the simulated data. This non-parametric technique guarantees a smooth representation of the distribution despite of the limited statistics\cite{kde}. 
 The green and blue vertical lines indicate the initial incident energies of the \ce{P} and F ions. Comparing these values to the mean values of the ion energy after crossing the a-\ce{SiO2} layer we conclude that on average \ce{P} ions lose 1 keV more energy than F ions in the amorphous layer on top of c-Si wafer. We also note that the energy spread of a mono-atomic \ce{P} ion after passing through the a-\ce{SiO2} layer is very similar, but yet the spread is slightly less broad with somewhat stronger pronounced maximum of the distribution. 

A molecular ion deposits three particles practically at the same time 
triggering simultaneous cascades in the target. One can consider hypothetically two extreme cases: at one extreme the three ions will evolve as separate mono-atomic ions producing an additive damage from the three independent cascades;
at the other, the molecule may keep as a whole until the full stop in the target material,
generating the damage as a single projectile  of the mass $M = M_{P} + 2 M_{F}$. In an intermediate scenario, the molecule splits in the impact but the three ions stay close until a certain depth triggering cascades in very close vicinity leading to cascade overlaps.

Assuming that during molecular ion irradiation, a molecular ion always breaks up into individual atoms at the moment when it hits the surface, one can use the BCA method to assess uncertainty of the \ce{P} atom placement at a given incident energy by simulating three independent cascades triggered by the mono-atomic particles of the molecule separately\cite{holmes2023improved}. Furthermore, the assumption is supported by the presence of natural amorphous oxide layer on the target surface, which presumably leads to spatial separation of the implanted particles that initially composed the molecular ion. To verify this hypothesis, we compare the results of MD simulations of implantation using mono-atomic and molecular ions. We carried out the simulations in both materials, pure c-\ce{Si} and a-\ce{SiO2}, to analyze the effect of the matrix on the separation rate of the atoms of the molecular ion. 

In these simulations, we tracked the distance between the atoms of the molecular ion along their paths inside the target. We followed the closest \ce{P} and F atoms until the distance between them did not exceed 1 nm in a PF complex or for all three atoms of \ce{PF2} molecule until the first of the two F atoms separated from the rest by the distance of 1 nm.  By recording the depth where the \ce{PF} complex and \ce{PF2} molecule broke up, we were able to estimate the percentage of the survived molecules and PF complexes as a function of depth. The decay of the percentage of survival with depth is plotted in Figure \ref{fig:effect_SiO2}b. We show that, for implantation in c-Si, in more than 40 \% of the cases, only one of the F atoms is detached from the molecule before 6 nm depth, while the second F atom keeps close to the \ce{P} atom. This fraction is reduced by the presence of the a-SiO$_{2}$ layer which likely increases the number of collisions.

The exemplary snapshots of the dynamic evolution of a cascade triggered by a molecular ion are shown in Figure \ref{fig:effect_SiO2}c. The molecule breaks at a shallow depth but only one of the three ions leaves the region where the cascade is evolving and can be treated separately. The cascades initiated by the remaining ions will evolve together increasing strongly the probability of the overlap. The synergistic damage effects resulted from the overlapping cascades are in the focus of the present study.

\begin{figure}[H]
    \centering
    \subfigure[]{\includegraphics[width=0.49 \textwidth]{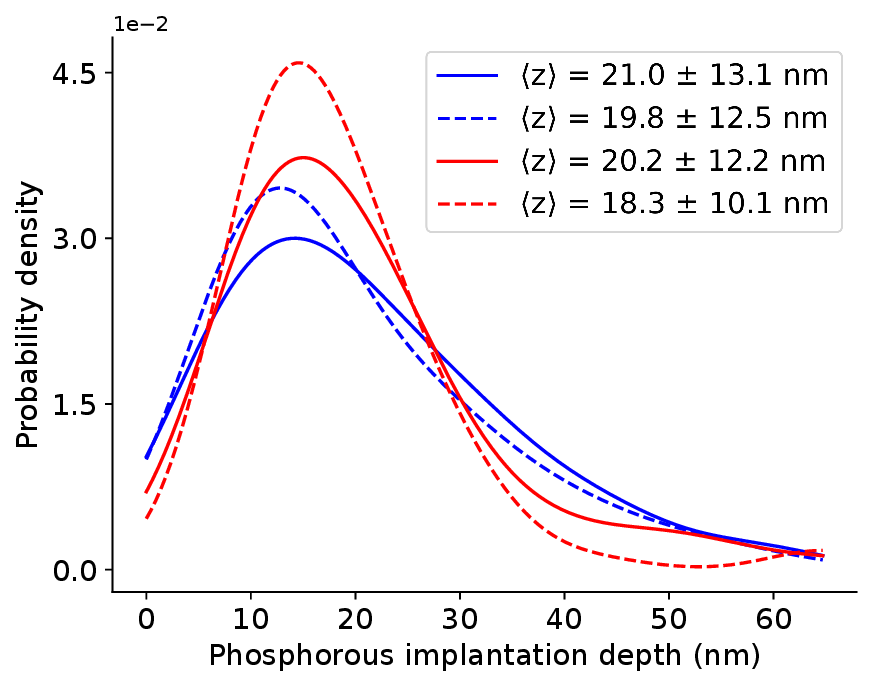}}
    \subfigure[]{\includegraphics[width=0.49 \textwidth]{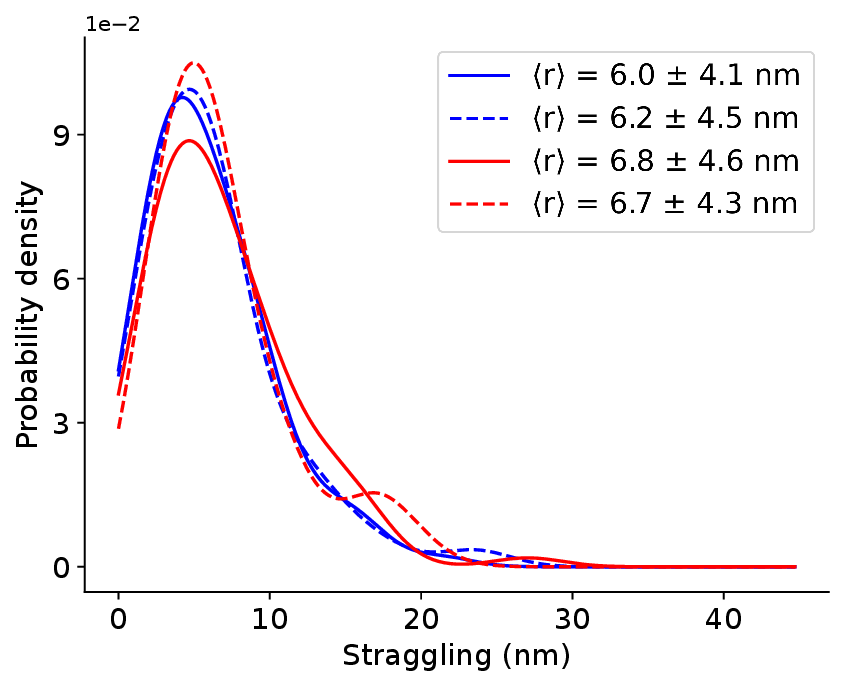}}
    \caption{P implantation profiles ((a) depth and (b) lateral) due to mono-atomic \ce{P}  (dashed) or a molecular \ce{PF2} (solid) ion implantation into pure c-Si (blue) or c-Si with an additional 5 nm a-\ce{SiO2} layer (red).}
    \label{fig:depth_profiles}
\end{figure}

Figure \ref{fig:depth_profiles}a shows the \ce{P} implantation profiles resulted from a mono-atomic (dashed lines) or from a \ce{PF2} molecular (solid lines) ion implantation. In this figure we see that the mono-atomic implantation of \ce{P} atoms in pure c-Si (blue colour in both panels of Figure \ref{fig:depth_profiles}) resulted in slightly sharper depth profile than the molecular implantation. The \ce{P} atoms are pushed deeper during the molecular ion irradiation and the overall depth profile is flatter, compared the profiles plotted in the dashed and solid lines in Figure \ref{fig:depth_profiles}a for mono-atomic and molecular ions, respectively. We see that, in many of the simulations, the implanted molecule does not split into individual atoms until a certain depth and, hence, acts similarly to a single particle of a larger mass promoting P ions deeper under the surface. When the molecule splits practically immediately, the P range is similar in both mono-atomic and molecular cases. This effect can explain why we obtained only slightly different shapes of the averaged P depth profiles for both mono-atomic and molecular ion irradiation. 

The effect is similar, although to lesser extent, when the \ce{P} atoms are  implantated in the c-Si substrate covered by the a-\ce{SiO2} layer (red color in both panels of Figure \ref{fig:depth_profiles}). Surprisingly, we see a stronger long tale formed by the channeled ions in case of the \ce{PF2} implantation, which we relate to higher energies of P atoms from a molecule, which they have when entering c-Si substrate after passing through the a-\ce{SiO2} layer, see Figure \ref{fig:effect_SiO2}a.

The spread in the lateral profiles in \ref{fig:depth_profiles}b is very similar for all four cases with a peak around 7 nm. However, when implanted as a mono-atomic ion, P lateral range exhibits a small peak in both substrates with and without the a-\ce{SiO2} top layer. This illustrates a stronger effect of the molecular ion irradiation on lateral channeling than that on the P channeling deeper into the bulk.

\subsection{Accuracy of placement detection based on electronic signal measurements}

\begin{figure}[H]
    \centering
    \subfigure[]{\includegraphics[width=0.49\textwidth]{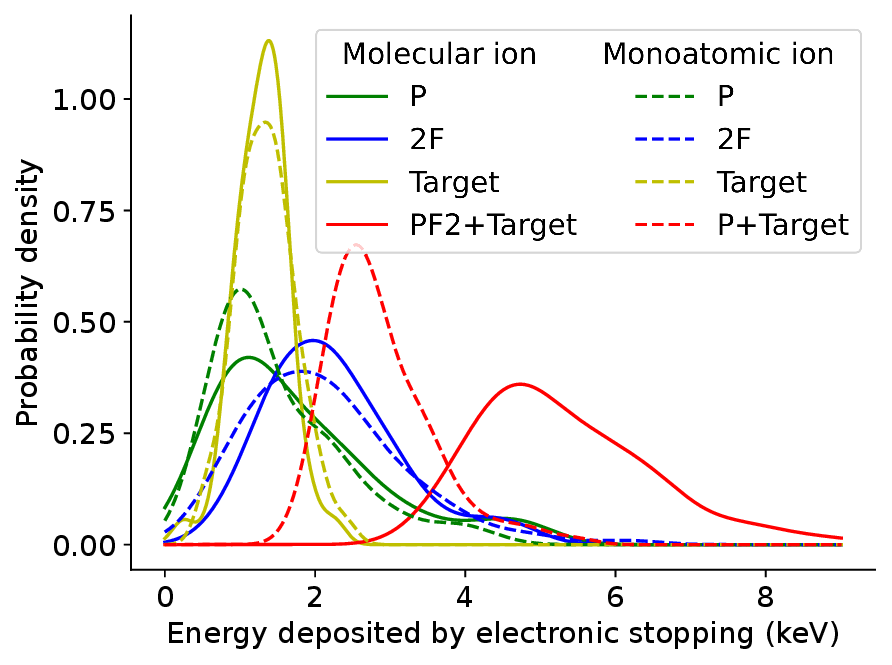}}
    \subfigure[]{\includegraphics[width=0.49\textwidth]{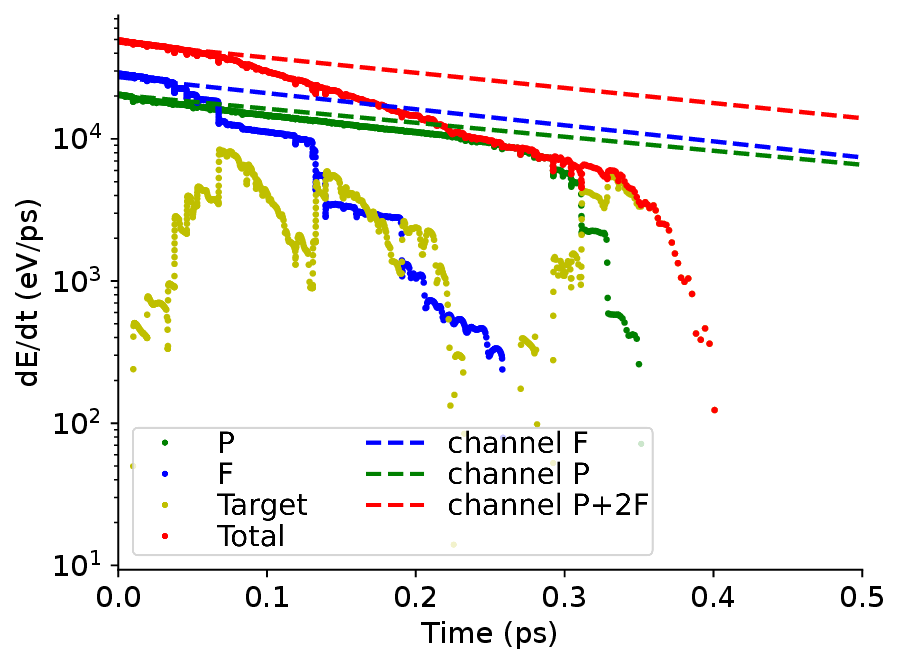}}
    \subfigure[]{\includegraphics[width=0.49\textwidth]{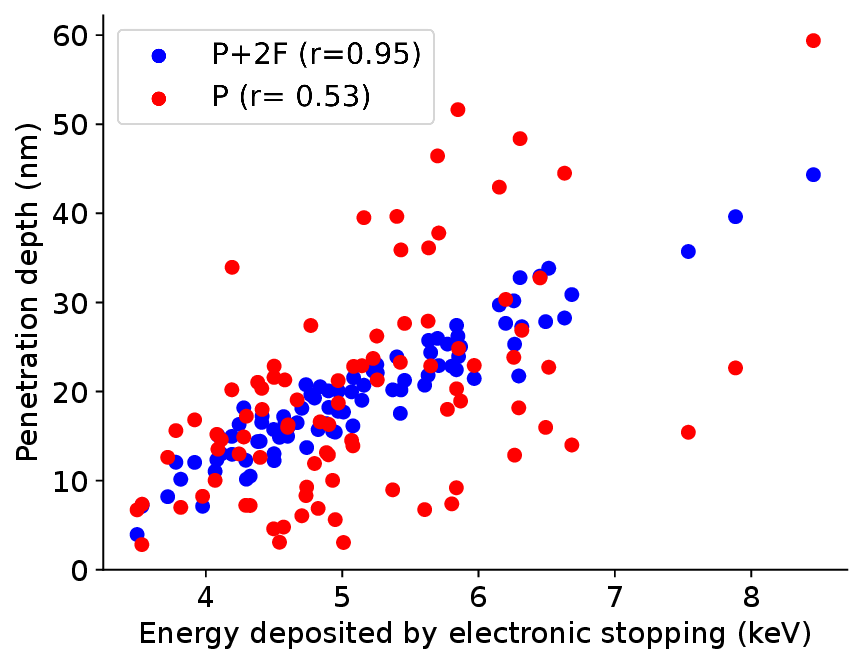}}
    \subfigure[]{\includegraphics[width=0.49\textwidth]{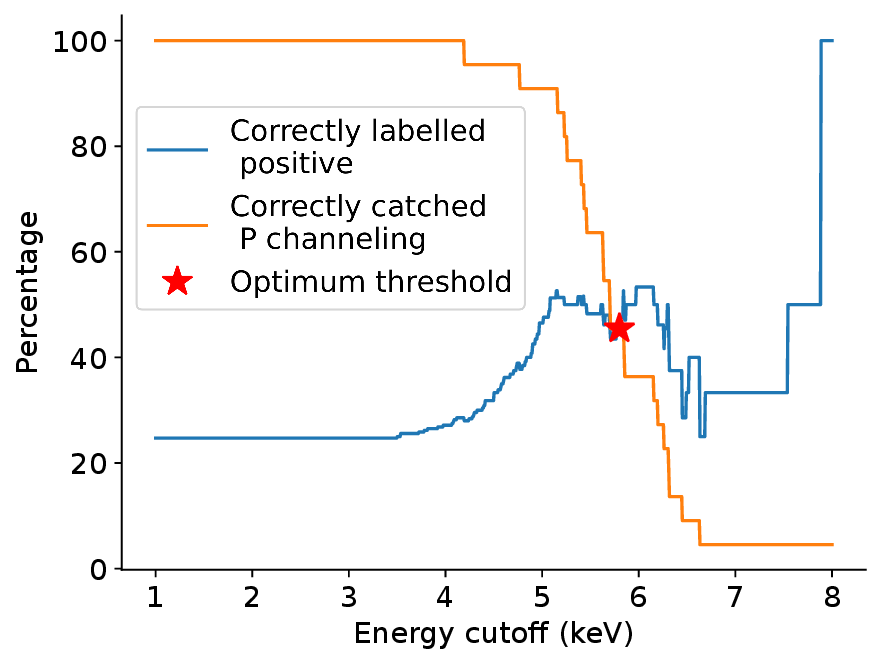}}
    \caption{Electronic stopping calculations recorded from simulations of mono-atomic ion and molecular ion implantation. The signal is used to determine the position of an implant. a) Total electronic stopping distributions for the molecule, its components and their sum. A KDE was fitted to the data. b) Time derivative for one implantation case and solution of the differential equation for the energy loss in the absence of nuclear collisions (dashed lines). c) Correlation plot of electronic stopping record and penetration depth of the ions. d) Performance of an hypothetical discriminator between channeling an non-channeling based on electronic stopping.}
    \label{fig:elstop_vs_penetration}
\end{figure}

Before analyzing the damage in c-\ce{Si} lattice due to molecular ion irradiation, we turn our attention to the electric measurements resulting from a single (mono-atomic or molecular) ion impact. When energetic particles interact with a material, they lose energy via two different mechanisms known as nuclear and electronic stopping. During the impact, some energy of the energetic particle is transferred to the electrons in the target atoms. As a result, the electrons become excited and with a certain probability an electron--hole pair is created. In turn, the latter is captured at the anode at the surface of the substrate in the form of an electronic signal, determining (or verifying) the position of the implanted atom. 

This method is straightforward for determination of the position of an implanted atom in an amorphous substrate when the former is implanted as a mono-atomic ion. However, mono-atomic ion implantation into a c-\ce{Si} may suffer from high channeling probability in the diamond lattice of silicon. Although the implantation is usually done with a small tilt to avoid channeling directions, the moving particles in the cascade can still be caught into channels increasing lateral and depth uncertainty of the implanted atom placement. 
Since channeling phenomenon constrains ion path within a single crystallographic direction, along its path in the channel the energetic particle experiences much fewer nuclear collisions compared to those in a random direction \cite{channeling}.
This implies that on average a channeled particle may be implanted deeper than a non-channeled one. Using 9 keV \ce{P} ion implantation, one can with high probability place the P atom very close to the desired depth of approximately 15 nm for a spin donor qubit, since the maximum of the range profile for 9 keV \ce{P} ions in silicon is also found at this depth, see Figure \ref{fig:depth_profiles}a. In the same figure one can see that the P range profile has a fairly long tail formed by channeled ions, which indicates a non-negligible probability of the implantation depth beyond 50 nm. Generally, the 7$^\circ$ tilt off normal can be used to suppress the channeling effect, however, one can take into account the presence of the natural a-\ce{SiO2} layer, which can randomize the incident angle of the ions before they enter the c-Si substrate. 
To investigate this effect, we performed the simulations of P implantation into c-Si after the ions passed through the 5 nm of a-\ce{SiO2} (similar to the experiment). Indeed in these simulations, the channeling fraction reduced from 25\% to 15\%. It is important to highlight that it was not possible to suppress the channeling completely and in some cases, the ions still reached far beyond the expected depth. 

In channeling condition, the electronic stopping becomes the dominant mechanism of the energetic particle deceleration in the material. This can be used for flagging channeling events. For instance, \cite{jakob2023scalable} suggests that the signal which determines the XY placement of \ce{P} atom could be further used to detect events of channeling of the implant.

In MD simulations, all the moving particles can be traced directly and the amount of energy deposited to the electronic subsystem can be recorded with the high precision.
In Figure \ref{fig:elstop_vs_penetration}a we plot the distribution of the electronic energy loss for all species moving in the target. Note that here we do not include 
the a-\ce{SiO2} layer in order to emphasize the channeling events. We plot this value for \ce{P} ions implanted using molecular (green solid line) or mono-atomic (green dashed line) ion irradiation. Similarly for other species we used solid lines for the molecular ion implantation and the dashed lines for the mono-atomic one. In the legend, "2F" refers to the signal from two F atoms, which were implanted by the \ce{PF2} molecule. For the mono-atomic implantation case, "2F" refers to the doubled signal from the simulations of a single F mono-atomic ion implantation. "Target" refers to the signal collected from the Si target atoms which were set in motion in cascades. "PF2+Target" and "P+Target" refer to the signals that can be measured experimentally during molecular and mono-atomic phosphorous implantation, respectively. With respect to \ce{P}, \ce{F} and Target signals we see that the largest difference in the shape of the energy loss distributions is observed for the \ce{P} and F atoms, with the stronger pronounced peak for mono-atomic \ce{P} ion. For both elements, P and F, the distributions are slightly shifted towards higher energies. Moreover, we also see a stronger "bump" in the high energy tail of the distributions, which is associated with longer paths of energetic particles in the lattice, i.e. with channeling. Meanwhile, the peaks for target atoms seem similar though slightly higher for the molecular ion. The signals shown in red are not comparable since the one corresponding to the molecular ion contains the contribution of 2 F atoms with starting energies of approximately 10 keV. However, those are the only signals that can be measured during the implantation of a \ce{PF2} molecule and a \ce{P} atom. We observe that the former signal is much stronger than the latter, but also in the lower range of energies, this is why the generated electric signal is less efficiently detected at the surface compared to the signal from a molecular ion. 

In Figure \ref{fig:elstop_vs_penetration}b we show a time evolution of the electronic stopping power for a single \ce{PF2} implantation run. Again "P", "F" and "Target" in the legend refer to the P and F projectiles and the Si secondary recoils generated in the cascade, respectively. The integral of the function labelled as "Total" (shown in red) yields the number of electrons constituting the measurable signal. Hence, it is important to understand how this integral is formed and how the shapes of the temporal evolution differ depending on whereas the ions are implanted deeper or shallower in the system. Here we chose to show the case with the channeling \ce{P} (green) projectile and non-channeling F (blue). We can observe that the curve from the \ce{P} ion decays slower than the one from F. 

For comparison, we also show the theoretical estimates of temporal evolution of electronic stopping power for a particle inside of a channel, see the dashed lines of the corresponding colour in Figure \ref{fig:elstop_vs_penetration}b calculated as follows: 
\begin{equation}
\frac{dE}{dx} = \underbrace{ F_{0}}_{0} - S_{e}(E) \Rightarrow  \frac{dE}{v dt} = - S_{e} (E) \Rightarrow \frac{dE}{dt} = - \sqrt{\frac{2E}{m}} S_{e}(E)
\end{equation}
Here we assume that electronic stopping acts as a friction force linearly decelerating the energetic particle moving in an electron cloud. Since we specifically selected \ce{P} atom to be trapped in a channel, where the number of nuclear collisions is very low and nuclear stopping contributes very little to particle deceleration. Therefore, the curve demonstrating $(\frac{dE}{dt})_{\text{e}}$ for the channeling \ce{P} atom is very close to the theoretical estimate (compare green curve with the green dashed line). 

The stopping power of the target material rises quickly from zero to the first maximum, but then fades somewhat before rising to the next peak. This behaviour is associated with ignition of large number of highly energetic recoils from the target atoms which also lose energy contributing to the total signal of the stopping power. Oscillations in this curve indicate the time separation between the subsequent strong impacts, in which a significant amount of energy was transferred to the recoils giving rise to the separated peaks in the temporal $(\frac{dE}{dt})_{\text{e}}$ curve. 

The contribution from channelled particles to the time evolution of the total electronic stopping power becomes more important at later stages of the ballistic phase, where the other particles have lost most of their energy and nuclear collisions prevail. As we see that after 0.25 ps the total stopping power practically coincides with the green curve, which shows the stopping power of the channeling P atoms. This could be used to discriminate between events of channeling. However, one can see that the signals generated by both species of a \ce{PF2} molecule,  \ce{P} and \ce{F}, in channeling direction are very similar (compare red and blue dashed lines showing theoretical estimates), hence, even if the time-dependent signal from the total electronic stopping could be measured, it would not be easy to differentiate between the signal from channeling \ce{P} or \ce{F} atoms. 

In experiments, the total electronic stopping is measured in terms of generated e-h pairs. In Figure \ref{fig:elstop_vs_penetration}c we correlate the penetration depth of \ce{P} atoms with the total electronic stopping, which can be measured in experiment. Each circle in the graph shows a different \ce{PF2} implantation case. The $x$ axis shows the energy of the signal and the $y$ axis is the penetration depth of the \ce{P} atoms during the implantation. Blue circles correlate the measured signal with the penetration depth calculated as an  average of the three components of the molecule, namely P and two F atoms. Red circles correlate the electronic stopping signal with the penetration depth of \ce{P} atoms only. Ideally, the circles would align along a straight line, corresponding to a direct mapping between the electronic signal intensity and the penetration depth of the \ce{P} atom. From this plot we observe that the correlation between the signal and the depth where the \ce{P} atoms have finally stopped is weak (the Pearson coefficient $r$, which is the measure of the similarity is about 0.5), while the average of the penetration depths of all three components correlate with the measured signal with the Pearson coefficient close to unity. 

We further explore the use of the signal as a \ce{P} implantation depth indicator. If we set the threshold for channeling events at implantation depths beyond 30 nm, an ideal discriminator would be able to catch 100\% of channeling events, while not labelling as positive any of non-channeling event. 

In machine-learning terms, we show in Figure \ref{fig:elstop_vs_penetration}d the true positive rate (TPR) and reverse of true negative rate (1-TNR) as a function of the signal value threshold beyond which the channeling events are safely excluded. Here by TPR corresponds to the probability of the correctly labelling the channeling events and TNR is the probability of labelling as "channeled" the cases, where the ions were not caught in a channel. Plotting the 1-TNR gives the percentage of correctly labelled channeling events. 

If the threshold is set too low, all implantation events are labelled as channeled. This results in 100\% of TPR, since all the channeled events were identified and none of them are lost. However, the 1-TNR is only 25\%, which exactly corresponds to the percentage of channeled ions in the simulations, hence, indicating that 75\% of implantation events were labelled as channeled wrongly. Increase of the threshold reduces the TPR value, but increases the 1-TNR value until these values cross and the 1-TNR value becomes very accurate as all detected events are labelled correctly as channeled. However, the reduction of TPR indicates that more and more events are not caught and many channeled events are included into proper implantation events.

We optimised the sum of the two fractions and found an appropriate threshold to be around 6 keV. For this parameter we see that we are unable to label properly around half of the events.  

Our results show that, when associating strong electronic signals to \ce{P} channeling events during molecular ion implantation, the uncertainty of the measurement may include also the similarity in the signals generated by different species. For instance, a detected strong signal could correspond to a channeling event of any of the three particles, not only of the P atoms. The positive side is that a low signal implies an event of non-channeling from any of the particles. Hence, one can 
safely discard events for which a strong signal was recorded, even if this may potentially discard also a correctly implanted P. 

\subsection{Damage analysis}

The ions, which are implanted in the substrate with the total energy of approximately 20 keV, decelerate not only via electronic, but also via intensive nuclear stopping. The latter generates energetic recoils, which trigger collision cascades displacing multiple atoms from their lattice positions. This forms stable defects which remain in the lattice even after the cascade and short-time post-cascade relaxation. 
It is critical for the practical application aimed in this study that there is no remaining damage in the lattice, and the phosphorous atom is in a substitutional position and electrically activated. In this section, we investigate the nature of stable damage formation during the energetic cascades triggered by the triple ions.

\begin{figure}[H]
    \centering
    \subfigure[]{\includegraphics[width=0.49\linewidth]{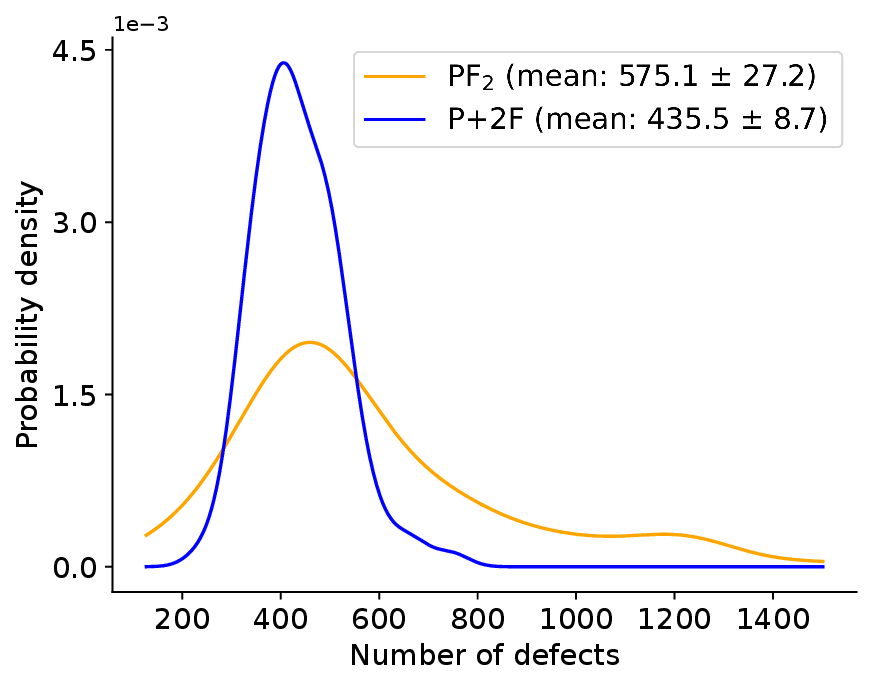}}
    \subfigure[]{\includegraphics[width=0.49\linewidth]{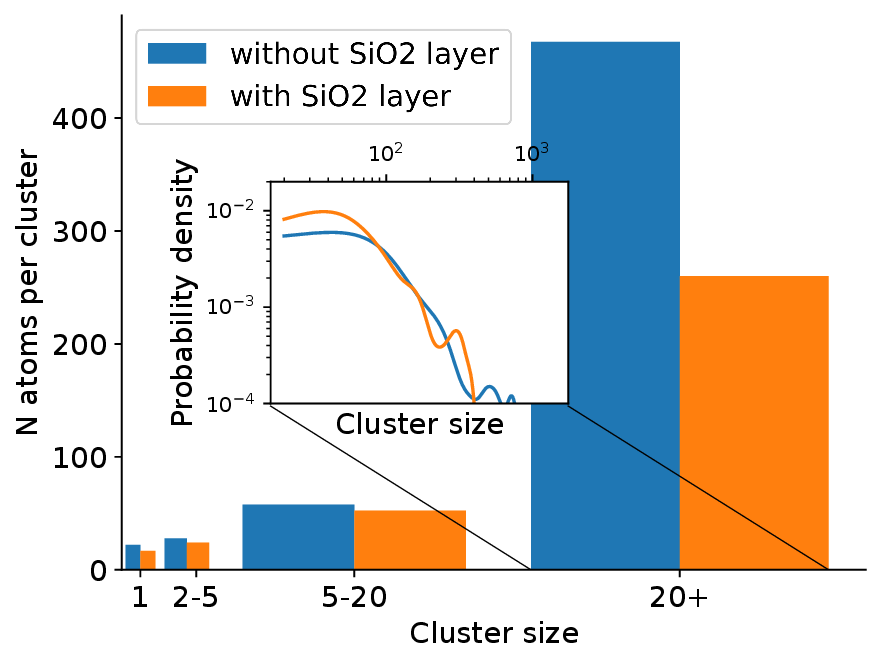}}
    \subfigure[]{\includegraphics[width=\linewidth]{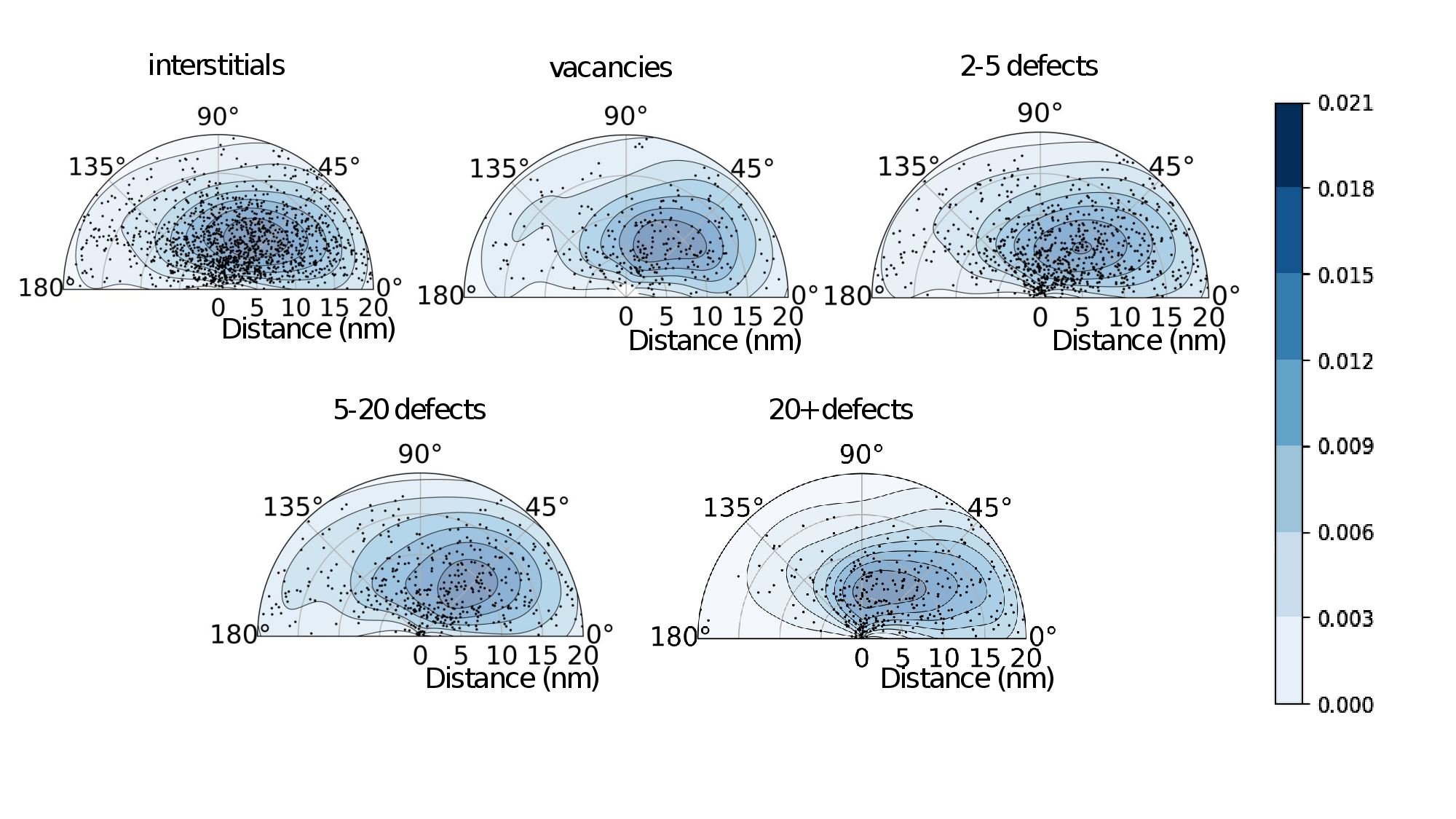}}
    \caption{a) Probability distribution obtained using a KDE of the number of defects obtained by wigner seitz analysis (vacancies + interstitials).The effect of the SiO$_{2}$ layer was included. b) For \ce{PF2} implantation, average per ion distribution of the damage in clusters scaled by the number of defects in the cluster. c) 2D scatter plot of the clusters center of mass distance and polar angle with respect to the \ce{P} atom position. A probability distribution was estimated by 2D KDE. a-SiO$_{2}$ layer was excluded}
    \label{fig:damage_Si}
\end{figure}

Figure \ref{fig:damage_Si}a shows the probability density distributions of the number of defects generated per a single implantation event. For comparability of the data, we plot 
in yellow 
the probability density distribution of defects 
generated by a molecular \ce{PF2} ion, while the blue line represents the distribution of the sum of the number of defects generated by a \ce{P} ions and two \ce{F} ions in a monoatomic ion irradiations. In addition we report the mean and the standard error of the mean for the total number of defects obtained per simulation. We observe that on average, while the total amount of energy introduced in the system is the same in both cases, the amount of defects generated by a molecular \ce{PF2} ion is somewhat greater than that produced jointly by the three constituent monoatomic ions with the same initial velocities as those of the corresponding particles within the molecular ion. The main difference is found in the spread of the distributions. The distribution of the number of defects produced jointly by the ions per implantation event generated jointly for three independent ion impacts is narrower (the standard deviation is below 100 defects) than that of the spread of the distribution for a molecular ion. The latter is fairly wide with the standard deviation as large as more than half of the mean value. 
This also means that the amount of defects which will be produced in a single ion implantation is more uncertain in case of molecular ion irradiation than that of a monoatomic ion irradiation. This result clearly indicates the high probability of internal overlaps of cascades triggered by each of the constituent atom of a \ce{PF2} molecule during the molecular ion implantation resulting in synergistic damage production compared to the three individual cases.

A large number of defects forming in a single ion implantation event, indicates that there is a high probability of defect clustering. Hence we analyzed the clusters of the defects produced in our simulations of ion impacts on a pure Si substrate during the \ce{PF2} ion irradiation by using a cluster analysis similar to what was done in \cite{clusters_analysis}.
In Figure \ref{fig:damage_Si}b we show the average number of point defects (1), small (2-5), medium (5-20 defects) and large (over 20 defects) clusters. We included the effect of the a-\ce{SiO2} top layer for comparison.
By examining the amount of actual defects inside the clusters, one can see that some of the clusters can reach the size 
over 500 atoms (see inset of Figure \ref{fig:damage_Si}b). From comparison of the results obtained with and without a protective a-\ce{SiO2} surface layer, 
we can see that the 
extra layer at the top of the substrate 
affects only the size of the largest clusters (over 20), while 
it does not change significantly the average size  of the smaller clusters. In the inset of figure \ref{fig:damage_Si}b we also see that the lower entry energy of the atoms from a molecular ion after it passed through the a-\ce{SiO2} layer (see Figure \ref{fig:depth_profiles}a) limits the size of the large cluster reducing to zero the probability of formation of very large clusters with the number of defects above 500. This behaviour may imply that the additional top surface layer will require lower annealing temperatures or shorter annealing times to recover the lattice structure around the implant. 

In Figure \ref{fig:damage_Si}c we analyze the radial and angular distributions of the defect clusters around the \ce{P} atom (the center of the radial distribution) for the clusters presented in Figure \ref{fig:damage_Si}b. In these graphs, the polar angle of $180^\circ$ points normally in the direction from the surface into the bulk. The distance between the \ce{P} atom and a cluster is measured as the distance from the location of the\ce{P} atom to the location of the center-of-mass of each given defect cluster. For clarity these distributions are shown only for pure Si substrate. To emphasize the difference in shape of interstitial and vacancy distributions, we plot them separately in two distinct plots.

First of all, we see point defects quite uniformly distributed around the implanted P atom. Although some point defects appear far from the implanted atom, the majority of these defects concentrate near it with the maximum of the distribution slightly behind the implanted atom itself. Interestingly the number of mono-interstitials is much higher than the number of mono-vacancies. Assuming the loss of atoms via sputtering is negligible as well as the vacancy migration at this early stage right after the cascade, the total number of vacancies must be equal or very close to that of the total number of interstitials. The dis-balance in number of vacancies and interstitials is explained by accumulation of the former in the damage clusters, which we confirmed by visual inspection of atomic displacements in individual cases. This implies that clusters of defects as well as amorphous pockets are vacancy rich, i.e. mass deficient. This may stabilize the formation of amorphous phase, since it provides the extended volume for formation less dense amorphous phase. In addition, \cite{holland1988damage} showed that this extended defects act as preferential interstitial sinks, favouring the appearance of di-vacancies. This increases the crystal's free energy, helping the amorphisation (see the review by \cite{Pelaz2004}).

The largest clusters (20+) again are distributed closer to the \ce{P} atoms, with the highest probability density within a few nanometres, see bottom right diagram in Figure \ref{fig:damage_Si}c. Since the average size of the large defect clusters is of the same size and larger, this result indicates that P atoms frequently belong within the large defect clusters as well, which are also likely to be amorphous pockets. 
This is expected since the large clusters can only be generated by the highest energies, and the phosphorous carries more than half of the energy of the molecule before the implantation. 
For the largest clusters, the damage tends to appear in the first quartile of the polar plot, therefore at shallower depths than P. In other words, the large cluster, i.e. the amorphous pocket, is generated during the intense slowing down of the \ce{P} atom before it finds itself in a final position.

Point defects in \ce{Si} have small migration activation energies \cite{ramanarayanan2003point} that make them mobile in the room-temperature irradiation conditions. Some of them may annihilate or form more stable small defect clusters which need high temperature annealing treatment to be removed from the system \cite[Chapter~5.5]{campbell2001science}. It has been shown that the amorphous pockets can be dissolved during high temperature annealing \cite{Rubia1996}. However, some small defect clusters and point defects, may survive as remnants of the amorphous pockets \cite{Rubia1995}. These remaining defects may induce strain in the lattice around the implanted P atom affecting quantum properties of the qubit. Additionally, point defects can pair with the P atom to form defects that are not suitable for the quantum applications. Examples of these are so-called E-centers formed by pairing of P atoms with the surrounding vacancies \cite{E_center}. These defects were shown to be unstable with the high probability of the vacancy jump away from the P atom. However, because of the specific migration path of the pair \cite{E_center}, this defect may affect the final location of the P atom. We also point out that some damage generated in our simulations will anneal or annihilate even before the thermal treatment is applied,  however, in the present study we show significant difference in the initial damage and its distribution in the condition used in the single ion implantation. This initial damage strongly correlates with the final remaining damage requiring additional treatment. The understanding of the nature of the initial damage formed by a molecular ion compared to that of the monoatomic ion can help in tuning the annealing temperature and annealing time. The control over these parameters is important with the purpose of suppressing detrimental effects of atomic diffusion at high temperatures during the long annealing times. 

\section{Conclusion}

In this work, by means of molecular dynamics simulations we provide a comprehensive study of the processes that take place during implantation of molecular \ce{PF2} ions compared to monoatomic \ce{P} ions. First, we examine the differences in the penetration depths, lateral stragglings, as well as channeling probability of the \ce{P} atoms introduced in monoatomic and molecular ion irradiation. Moreover, we assess the effect of an a-\ce{SiO2} layer on top of the Si wafer on the studied quantities. Next, we show that while the a-\ce{SiO2} layer attenuates the collaborative effect of the simultaneous introduction of three ions on the probability of channeling, the latter events still take place during the implantation, 
and thus need to be detected. However, we demonstrate that when using a molecule instead of a monoatomic ion, a high signal fro, electronic energy loses 
may not be produced by a deep penetration of the \ce{P} atom but by any of the three atoms in the \ce{PF2} molecule. Therefore, one needs to be cautious when associating high signal to \ce{P} channeling events. Finally, we have analyzed the damage caused by the ions in the lattice, which could be problematic for the applications, if not recovered properly. We demonstrate that the collaborative effect of the \ce{PF2} adds more defects to the system than if the \ce{P} ions were implanted separately, and consequently these defects group in large clusters. These clusters are rather centered around the \ce{P} atom. They take longer times to anneal compared to point defects and could give rise to undesirable structures. 

\section*{Disclosure statement}
No potential conflict of interest was reported by the author(s).

\section*{Acnoweledgements}
We gratefully acknowledge the Academy of Finland, SPATEC
project (grant nº 1349690) for financial support and the IT Centre of Science CSC
for providing CPU resources. Additionally, we gratefully acknowledge excellent discussions with Prof. David Jamieson and Dr
Malwin Jakob from the university of Melbourne. 

\section*{Data availability}
The data that support the findings of this study are available from the corresponding author, [author initials], upon reasonable request.

\bibliographystyle{vancouver} 
\bibliography{refs}
\end{document}